\documentclass[
twocolumn,
aps,
prl,
reprint,
superscriptaddress,
amsmath,amssymb
]{revtex4-1}
\usepackage[pdftex]{graphicx} 
\usepackage[hidelinks]{hyperref} 
\hypersetup{
 setpagesize=false,
 bookmarksnumbered=true,%
 bookmarksopen=true,%
 colorlinks=true,%
 linkcolor=blue,
 citecolor=blue,
}

\usepackage{braket}
\usepackage{times,multirow,amsfonts,bm,xspace,pifont}
\usepackage{soul}
\usepackage{ulem}
\usepackage{mathrsfs}
\begin{document}
\newcommand{\rr}{{\bm r}}
\newcommand{\q}{{\bm q}}
\renewcommand{\k}{{\bm k}}
\newcommand*\wien    {\textsc{wien}2k\xspace}
\newcommand*\textred[1]{\textcolor{red}{#1}}
\newcommand*\textblue[1]{\textcolor{blue}{#1}}
\newcommand*\YY[1]{\textcolor{blue}{#1}}
\newcommand{\YYS}[1]{\textcolor{blue}{\sout{#1}}}
\newcommand*\JI[1]{\textcolor{red}{#1}}
\newcommand*\JIS[1]{\textcolor{red}{\sout{#1}}}
\newcommand*\AD[1]{\textcolor{magenta}{#1}}
\newcommand*\ADS[1]{\textcolor{magenta}{\sout{#1}}}
\newcommand*\ADC[1]{\textcolor{green}{#1}}
\newcommand{\TK}[1]{{\color{red}{#1}}}
\newcommand\nazo[1]{\textcolor{red}{#1}}

\title{Superconductivity in monolayer FeSe enhanced by quantum geometry
}

\author{Taisei Kitamura}
\email[]{kitamura.taisei.67m@st.kyoto-u.ac.jp}
\affiliation{Department of Physics, Graduate School of Science, Kyoto University, Kyoto 606-8502, Japan}

\author{Tatsuya Yamashita}
\affiliation{Department of Physics, Graduate School of Science, Kyoto University, Kyoto 606-8502, Japan}

\author{Jun Ishizuka}
\affiliation{Institute for theoretical physics, ETH Zurich, 8093 Zurich, Switzerland}

\author{Akito Daido}
\affiliation{Department of Physics, Graduate School of Science, Kyoto University, Kyoto 606-8502, Japan}

\author{Youichi Yanase}
\affiliation{Department of Physics, Graduate School of Science, Kyoto University, Kyoto 606-8502, Japan}
\affiliation{%
  Institute for Molecular Science, Okazaki 444-8585, Japan
}
\date{\today}

\begin{abstract}
We formulate the superfluid weight in unconventional superconductors with $\bm k$-dependent Cooper pair potentials based on the geometric properties of Bloch electrons. 
We apply the formula to a model of the monolayer FeSe obtained by the first-principles calculation.
Our numerical calculations point to a significant enhancement of the Berezinskii-Kosterlitz-Thouless transition temperature 
due to the geometric contribution to the superfluid weight, which is not included in the Fermi liquid theory.
The $\bm k$-dependence of the gap function also stabilizes the superconducting state.
Our results reveal that the geometric properties of Bloch electrons play an essential role in superconducting materials 
and pave the way for clarifying hidden aspects of superconductivity from the viewpoint of quantum geometry. 
\end{abstract}

\maketitle


Monolayer FeSe grown on SrTiO$_3$ has been reported to experience a superconducting transition at the transition temperature $T_{\rm c}$ higher than $65$ K~\cite{wang2012interface-induced,he2013phase,xu2020spectroscopic}, in stark contrast to $T_{\rm c}\sim 8$ K of the bulk FeSe~\cite{hsu2008superconductivity}.
Such a significant enhancement of the transition temperature has been attracting much attention, but its origin remains to be clarified.
In two-dimensional superconductors, the resistive transition is determined by the Berezinskii-Kosterlitz-Thouless (BKT) transition~\cite{kosterlitz1973ordering,berezinskii1971destruction},
where the BKT transition temperature $T_{\rm BKT}$ is given by the superfluid weight $D^{\rm s}(T)$ according to the formula $D^{\rm s}(T_{\rm BKT}) = 8T_{\rm BKT}/\pi$~\cite{beasley1979possibility,gao2014gapless,williams1998vortex}.
Thus, the study of the superfluid weight is the key to understanding the significant enhancement of the transition temperature in monolayer FeSe. It should also be noted that the superfluid weight is essential for the Meissner effect and is related to the magnetic penetration depth by $\lambda(T)=1/\sqrt{4\pi D^{\rm s}(T)}$.
Considering the temperature dependence of $D^{\rm s}(T)$ and $\lambda(T)$ is closely related to the gap structure
, the superfluid weight is a useful probe of the pairing symmetry \cite{biswas2018direct, yao2019diamagnetic, khasanov2010evolution, takahashi2011anomalous, biswas2018evidence}.
Thus, the evaluation of the superfluid weight is essential to explore the nature of the high-temperature superconductivity in monolayer FeSe.

In the Fermi-liquid theory, the superfluid weight has been believed to be determined by the density of electrons $n^*$ and the effective mass $m^*$, i.e. $n^*/m^*$~\cite{jujo2001fermi,tinkham2004introduction,comment}.
However, recent studies have revealed that the geometric properties of the Bloch electrons can contribute to the superfluid weight in multi-band superconductors~\cite{peotta2015superfluidity,liang2017band}.
When the bands are sufficiently apart from each other, this contribution reduces to the quantum metric of the Bloch wave function~\cite{liang2017band}.

The quantum metric is closely related to the Berry curvature through the quantum geometric tensor~\cite{resta2011the,provost1980riemannian}.
The imaginary part of the quantum geometric tensor is widely known as the Berry curvature~\cite{berry1984quantal}, which appears in various Hall responses~\cite{thouless1982quantized,xiao2010berry,Nagaosa2010}.
The real part is the quantum metric, which 
also appears in physical properties of solids,
such as spread of the Wannier functions~\cite{marzari1997maximally}, current noise~\cite{Neupert2013}, exciton energy levels~\cite{Srivastava2015}, 
positional shift by an external electric field~\cite{gao2014field} and the electric quadrupole moments~\cite{gao2019nonreciprocal,lapa2019semiclassical,daido2020thermodynamic}.
The quantum metric can be divided into the contribution from each band, which is especially called the band-resolved quantum metric.
The band-resolved quantum metric has been revealed to be an essential ingredient of nonlinear optical responses such as photocurrent generation~\cite{ahn2020low-frequency,watanabe2021chiral}.
Thus, the geometric properties of the Bloch electrons play an essential role in understanding the material properties.

The importance of the geometric contribution to the superfluid weight was discussed for the first time in the Lieb optical lattice of cold atoms~\cite{julku2016geometric,he2021geometry}, where the flat band has been realized~\cite{taie2015coherent,ozawa2017interaction-driven}.
In the flat-band limit, the conventional contribution disappears, because $m^*\rightarrow\infty$ leads to $n^{*}/m^*\rightarrow0$; this implies the dominant geometric contribution.
Furthermore, in the newly discovered superconducting twisted bilayer graphene~\cite{cao2018unconventional}, in which the moire flat-band appears~\cite{li2010observation,bistritzer2011moire},
the geometric contribution is shown to be dominant in the superfluid weight ~\cite{hu2019geometric,julku2020superfluid,xie2020topology-bonded,wang2020quantum}.
Thus, an essential role of the geometric properties of the Bloch electrons has been recognized for the superconductivity in artificial quantum systems.

In this Letter, 
we show that FeSe manifests the geometric contribution to the superfluid weight without artificial electronic structure.
Because FeSe is a multi-band superconductor and the mother compound of a topological superconductor candidate FeSe$_{1-x}$Te$_x$~\cite{wang2015topological,xu2016topological,wang2018evidence,zhang2018obsevation,machida2019zero-energy}, the geometric properties of the Bloch electrons should be nontrivial and may cause intriguing phenomena.
Actually, four of the authors have shown that 
the geometric properties of Bloch electrons lead to a finite electric quadrupole moment~\cite{kitamura2021thermodynamic}.
Furthermore, 
FeSe has a small carrier density $n^*$, and thus in the Bardeen-Cooper-Schrieffer to Bose-Einstein-Condensation (BCS-BEC) crossover regime~\cite{nozieres1985bose,kasahara2014field-induced,kasahara2016giant,hanaguri2019quantum,kasahara2020evidence}.
Therefore, the geometric contribution to the superfluid weight is naturally expected to have a significant effect on the superconducting FeSe.
Thus, FeSe may offer a promising platform to study the geometric effects in non-artificial superconductors.

\textit{Formulation of superfluid weight.} ---
In the previous study based on the BCS theory~\cite{liang2017band}, the superfluid weight with the $\bm k$-independent pairing is divided into two terms: One is the conventional term while the other is the geometric term.
We extend the formulation to describe unconventional superconductivity. 
The superfluid weight will be divided into four terms as will be shown later in Eqs.~\eqref{sfw_all}-\eqref{gap}.

We start from the Bogoliubov-de-Gennes (BdG) Hamiltonian, 
$
\hat{H}_{\rm BdG} = \sum_{\bm k}\hat{\psi}^\dagger_{\bm k}H_{{\rm BdG}\bm k}\hat{\psi}_{\bm k},
$ with
\begin{eqnarray}
  &H_{{\rm BdG} \bm k} =
  \left(
  \begin{array}{cc}
    H_{0\bm k}&\bm \Delta_{\bm k}\\
    \bm \Delta^\dagger_{\bm k}&-H_{0-\bm k}^T
  \end{array}
  \right)&,
\end{eqnarray}
where $\bm k$ is the wave vector and $\hat{\psi}_{\bm k}$ is the Nambu spinor written by
$
  \hat{\psi}_{\bm k} = \left(
    \hat{c}_{1\uparrow\bm k},  \cdots  ,\hat{c}_{f\uparrow\bm k}, \hat{c}^\dagger_{1\downarrow\bm -\bm k},  \cdots  ,\hat{c}^\dagger_{f\downarrow\bm -\bm k}
  \right)^{T}.
$
Here, $\hat{c}^\dagger_{i\sigma\bm{k}}$ ($\hat{c}_{i\sigma\bm{k}})$ is the creation (annihilation) operator, 
$i=1,2,\cdots f$ shows the orbital and sublattice indices, and $\sigma=\uparrow,\downarrow$ represents the spin. We denote by $f$ the total number of orbital and sublattice degrees of freedom.
We ignore the spin-orbit coupling and assume spin-singlet superconductivity with iron-based superconductors in mind.
$H_{0\bm{k}}$ and $\bm \Delta_{\bm k}$ are the matrix representation of the Fourier transform of hopping integrals and the gap function, respectively.

The current response of superconductors to the vector potential $A_\mu(\bm q,\omega)$ is described by the Meissner kernel $K_{\mu\nu}(\bm{q},\omega)$, as
$
  j_\mu(\bm q,\omega) = -K_{\mu\nu}(\bm q, \omega)A_\nu(\bm q, \omega).
$
The superfluid weight $D^{\rm s}_{\mu\nu}$ is defined by its $q$ limit,
$
  D^{\rm s}_{\mu\nu} = \lim_{\bm q\rightarrow0}K_{\mu\nu}(\bm q,0).
$
According to the Kubo formula, the superfluid weight is obtained as
\begin{eqnarray}
  D^{\rm s}_{\mu\nu} &=& \sum_{\bm k\alpha\beta}\dfrac{f(E_{\alpha\bm k})-f(E_{\beta\bm k})}{E_{\alpha\bm k}-E_{\beta\bm k}}\notag\\
  &\times&
  (
  \bra{\psi_{\alpha\bm k}}\partial_{\mu}H_{p\bm k}\ket{\psi_{\beta\bm k}}\bra{\psi_{\beta\bm k}}\partial_{\nu}H_{p\bm k}\ket{\psi_{\alpha\bm k}}\notag\\
  &-&\bra{\psi_{\alpha\bm k}}\partial_{\mu}H_{{\rm BdG} \bm k}\ket{\psi_{\beta\bm k}}\bra{\psi_{\beta\bm k}}\partial_{\nu}H_{m\bm k}\ket{\psi_{\alpha\bm k}}
  ), 
\end{eqnarray}
where we introduced block-diagonal matrices
\begin{eqnarray}
  H_{p(m) \bm k} =
  \left(
  \begin{array}{cc}
    H_{0\bm k}&0\\
    0&(-)H_{0-\bm k}^T
  \end{array}
  \right).
\end{eqnarray}
The wave function and the energy eigenvalue of the BdG Hamiltonian are denoted by $H_{\mathrm{BdG}\bm k}\ket{\psi_{\alpha\bm k}}=E_{\alpha\bm k}\ket{\psi_{\alpha\bm k}}$.

To classify the superfluid weight by the geometric properties of the normal state, we introduce the energy and the Bloch wave function, i.e.
$H_{0\bm k}\ket{u_{n\bm k}} = \epsilon_{n\bm k}\ket{u_{n\bm k}}$, following Ref.~\cite{liang2017band}.
For simplicity, we assume the time-reversal symmetry, 
under which $H_{0\bm k} = H_{0-\bm k}^T$ is satisfied.
Using the matrix elements $\phi_{n\bm k}^{i\uparrow(\downarrow)}$ of the unitary matrix which diagonalizes the BdG Hamiltonian, 
the wave function of the BdG Hamiltonian is expanded by the normal state Bloch wave function as,
\begin{eqnarray}
  \ket{\psi_{\alpha\bm k}} = \left(
  \begin{array}{c}
    \sum_n\phi_{n\bm k}^{\alpha\uparrow}\ket{u_{n\bm k}}\\
    \sum_n\phi_{n\bm k}^{\alpha\downarrow}\ket{u_{n\bm k}}
  \end{array}
  \right).
\end{eqnarray}
By using this relationship, the superfluid weight for unconventional superconductors is divided into four parts as follows:
\begin{eqnarray}
  &&D^{\rm s}_{\mu\nu} = D^{\rm conv}_{\mu\nu} + D^{\rm geom}_{\mu\nu} + D^{\rm multi}_{\mu\nu} + D^{\rm gap}_{\mu\nu},\label{sfw_all}\\
  &&D^{\rm conv}_{\mu\nu} = 2\sum_{nm\bm k}C_{nnmm\bm k}^{\uparrow\uparrow\downarrow\downarrow}(J_{nn\bm k}^{\mu}J_{mm\bm k}^{\nu}+J_{nn\bm k}^{\nu}J_{mm\bm k}^{\mu}),\label{conv}\\
  &&D^{\rm geom}_{\mu\nu} = 2\sum_{n\neq m, l\neq s\bm k}C_{nmls\bm k}^{\uparrow\uparrow\downarrow\downarrow}(J_{nm\bm k}^{\mu}J_{ls\bm k}^{\nu}+J_{nm\bm k}^{\nu}J_{ls\bm k}^{\mu}),\label{geom}\\
  &&D^{\rm multi}_{\mu\nu} = 2\sum_{nl\neq s\bm k}\left(C_{nnls\bm k}^{\uparrow\uparrow\downarrow\downarrow}\right.
  \left(J_{nn\bm k}^{\mu}J_{ls\bm k}^{\nu}+J_{nn\bm k}^{\nu}J_{ls\bm k}^{\mu}\right)\notag\\
  &&
\left. +C_{lsnn\bm k}^{\uparrow\uparrow\downarrow\downarrow}
  \left(J_{ls\bm k}^{\mu}J_{nn\bm k}^{\nu}+J_{ls\bm k}^{\nu}J_{nn\bm k}^{\mu}\right)\right),\\
  &&D^{\rm gap}_{\mu\nu} =
  \sum_{nmls\sigma\bm k}
  S\left(C_{nmls\bm k}^{\uparrow\downarrow\sigma\sigma}\delta\Delta_{nm\bm k}^{\mu}+C_{nmls\bm k}^{\downarrow\uparrow\sigma\sigma}\delta\Delta_{nm\bm k}^{\dagger\mu}\right)
  J_{ls\bm k}^{\nu},\label{gap}\notag\\
\end{eqnarray}
where $S$ takes $-(+)$ when $\sigma = \uparrow(\downarrow)$.
Here, 
$J_{nm\bm k}^{\mu}$, $\delta\Delta_{nm\bm k}^{\mu}$, $\delta\Delta_{nm\bm k}^{\dagger\mu}$,
and 
$C_{nmls\bm k}^{\sigma_1\sigma_2\sigma_3\sigma_4}$ 
are written by
\begin{eqnarray}
  &&J_{nm\bm k}^{\mu} = \bra{u_{n\bm k}}\partial_\mu H_{0\bm{k}}\ket{u_{m\bm k}},\\
  &&\delta\Delta_{nm\bm k}^{\mu} = \bra{u_{n\bm k}}\partial_\mu \bm\Delta_{\bm k}\ket{u_{m\bm k}},\\
  &&\delta\Delta_{nm\bm k}^{\dagger\mu} = \bra{u_{n\bm k}}\partial_\mu \bm\Delta^\dagger_{\bm k}\ket{u_{m\bm k}},\\
  &&C_{nmls\bm k}^{\sigma_1\sigma_2\sigma_3\sigma_4} = \sum_{\alpha\beta\bm k}\dfrac{f(E_{\alpha\bm k})-f(E_{\beta\bm k})}{E_{\alpha\bm k}-E_{\beta\bm k}}
  \phi_{n\bm k}^{\alpha\sigma_1*}\phi_{m\bm k}^{\beta\sigma_2}\phi_{l\bm k}^{\beta\sigma_3*}\phi_{s\bm k}^{\alpha\sigma_4}.\notag\\
\end{eqnarray}
The details of the derivation are shown in Supplemental Materials~\cite{Supplemental}.

The conventional term $D^{\rm conv}_{\mu\nu}$ is found in the first term of Eq.~\eqref{sfw_all}.
Since $J_{nn\bm k}^{\mu} = \partial_\mu\epsilon_{n{\bm k}}$, 
this term is essentially determined by the energy dispersion.
Only this term is studied in the Fermi-liquid theory.
The second term $D^{\rm geom}_{\mu\nu}$ of Eq.~\eqref{sfw_all} is the interband effect. 
This term is called the geometric term, as
the interband velocity operator appears 
in Eq.~\eqref{geom}, $J_{nm\bm k}^{\mu} = (\epsilon_m-\epsilon_n)\braket{u_{n\bm k}\vert\partial_\mu u_{m\bm k}}$, which represents the geometric properties of the Bloch wave function.
In the absence of the interband pairing, the terms with $n\neq l, m\neq s$ vanish, and Eq.~\eqref{geom} is represented by the band-resolved quantum metric~\cite{Supplemental}.
The third term $D^{\rm multi}_{\mu\nu}$, called the multi-gap term, vanishes in the case of band-independent pairing.
We show that this term is negligible in monolayer FeSe later.

The fourth term $D^{\rm gap}_{\mu\nu}$ of Eq.~\eqref{sfw_all} comes from the $\bm k$-dependence of the gap function and directly reflects the pairing symmetry.
We call this term the gap term.
This term has been neglected in the previous studies~\cite{julku2020superfluid}.
However, various pairing states, such as $s_{++}$-wave, 
nodeless $d$-wave, and incipient $s_{\pm}$-wave states, have been suggested for monolayer FeSe~\cite{chen2015electron,gao2016hidden,kang2016superconductivity,yamakawa2017superconductivity,agterberg2017resilient,huang2017monolayer,schrodi2020multichannel}, and some proposals assume ${\bm k}$-dependent gap functions.
Furthermore, this term is required to reproduce the conventional formula, $D \propto n^*/m^*$,
when $\bm \Delta_{\bm k} 
= \bm 1 \times \Delta_{\bm k}$ 
(see Supplemental Materials~\cite{Supplemental} for details).

\textit{10-orbital Model.} ---
To calculate the superfluid weight of monolayer FeSe, we construct a realistic 10-orbital tight-binding model for Fe $3d$ orbitals.
The first-principles electronic structure calculation is performed using the \wien code~\cite{blaha2019wien2k}, and the tight-binding models based on the maximally localized Wannier functions~\cite{marzari1997maximally,souza2001maximally} are constructed by the WANNIER90 code~\cite{mostofi2008wannier90}.
The presence of two iron atoms in the unit cell doubles the number of orbitals, as $2\times 5 = 10$.

First, we construct a model of the bulk FeSe.
The results of angle-resolved photoemission spectroscopy 
are known to be slightly different from first-principles calculations~\cite{maletz2014unusual,zhang2016distinctive}. 
To reproduce the experimentally observed Fermi surfaces, we take into account an additional hopping parameter (see Supplemental Materials~\cite{Supplemental}).


\begin{figure}[tbp]
  \includegraphics[width=1.0\linewidth]{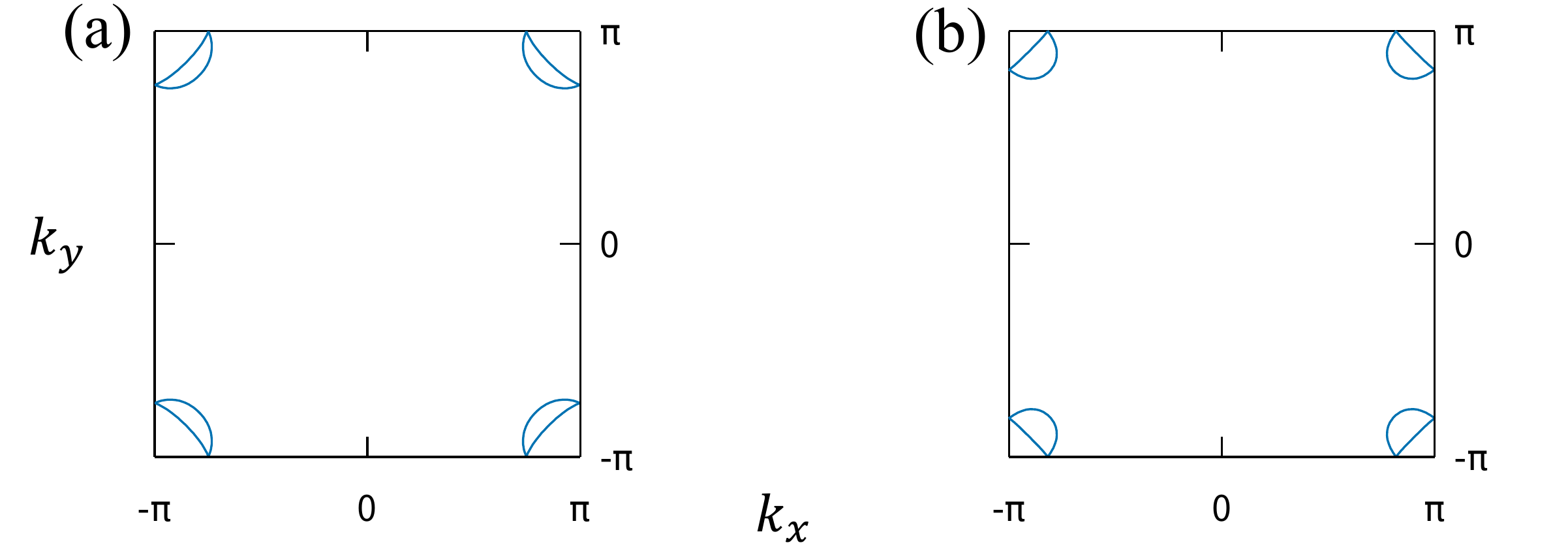}
  \centering
  \caption{Fermi surfaces of the model for monolayer FeSe. 
  (a) $n = 6.1$ and (b) $6.06$.
  The hole-like Fermi surfaces disappear in all the cases.
  \label{fig:fs_fese_2}}
\end{figure}

In monolayer FeSe grown on SrTiO$_3$, the hole-like Fermi surfaces at $\Gamma$ point disappear owing to the excess electron doping~\cite{miyata2015high-temperature},
which can be reproduced by shifting the chemical potential.
In this Letter, we consider three values for the particle number, $n = 6.1$, $6.08$, and $6.06$, corresponding to the excess electron doping $n-6$. 
The electron doping can also be caused by the K doping~\cite{miyata2015high-temperature} and the gate voltage~\cite{hanzawa2016electric,Shiogai2016}.
In addition, taking into account the mass enhancement by the electron correlation~\cite{aichhorn2010theoretical,yin2011kinetic,maletz2014unusual}, we renormalize the normal-state Hamiltonian as $z \hat{H}_0$ instead of the bare one $\hat{H}_0$, with choosing $z = 1/5$ or $1/8$.
It is known that the conventional term of the superfluid weight is renormalized by $z$ while the geometric terms are hardly affected.
Therefore, the renormalization effect may be essential for the origin of the superfluid weight.
The Fermi surfaces of the models are shown in Fig.~\ref{fig:fs_fese_2}.

\textit{Superfluid weight in monolayer FeSe.} ---
In the 10-orbital tight-binding model, the gap functions 
may be orbital and ${\bm k}$ dependent, making the multi-gap and gap terms finite. 
To determine the gap function and the mean-field transition temperature $T_{\rm c}$, we solve the gap equation
$
\Delta_{ij\bm k} = \sum_{\bm k^\prime} V_{ij\bm k\bm k^\prime} \braket{\hat{c}_{j\downarrow-\bm k^\prime} \hat{c}_{i\uparrow\bm k^\prime}}
$ self-consistently, with
$V_{ij\bm k\bm k^\prime}$ the pairing interaction.
We examine some candidates for the pairing state in monolayer FeSe, namely, the $s_{++}$-wave and incipient $s_{\pm}$-wave states. Furthermore, we study the $\bm k$-independent gap function as well for comparison.
For all cases, we phenomenologically determine the attractive interaction 
$V_{ij\bm k\bm k^\prime}$
so as to reproduce $T_{\rm c} \approx 83$~K, because the mean-field transition temperature of monolayer FeSe on SrTiO$_3$ is considered to be among $65$~K to $83$~K~\cite{wang2012interface-induced,he2013phase,xu2020spectroscopic}.
Similar results are obtained for $T_{\rm c} = 65$~K as shown in Supplemental Materials~\cite{Supplemental}.

\begin{figure}[tbp]
  \includegraphics[width=1.0\linewidth]{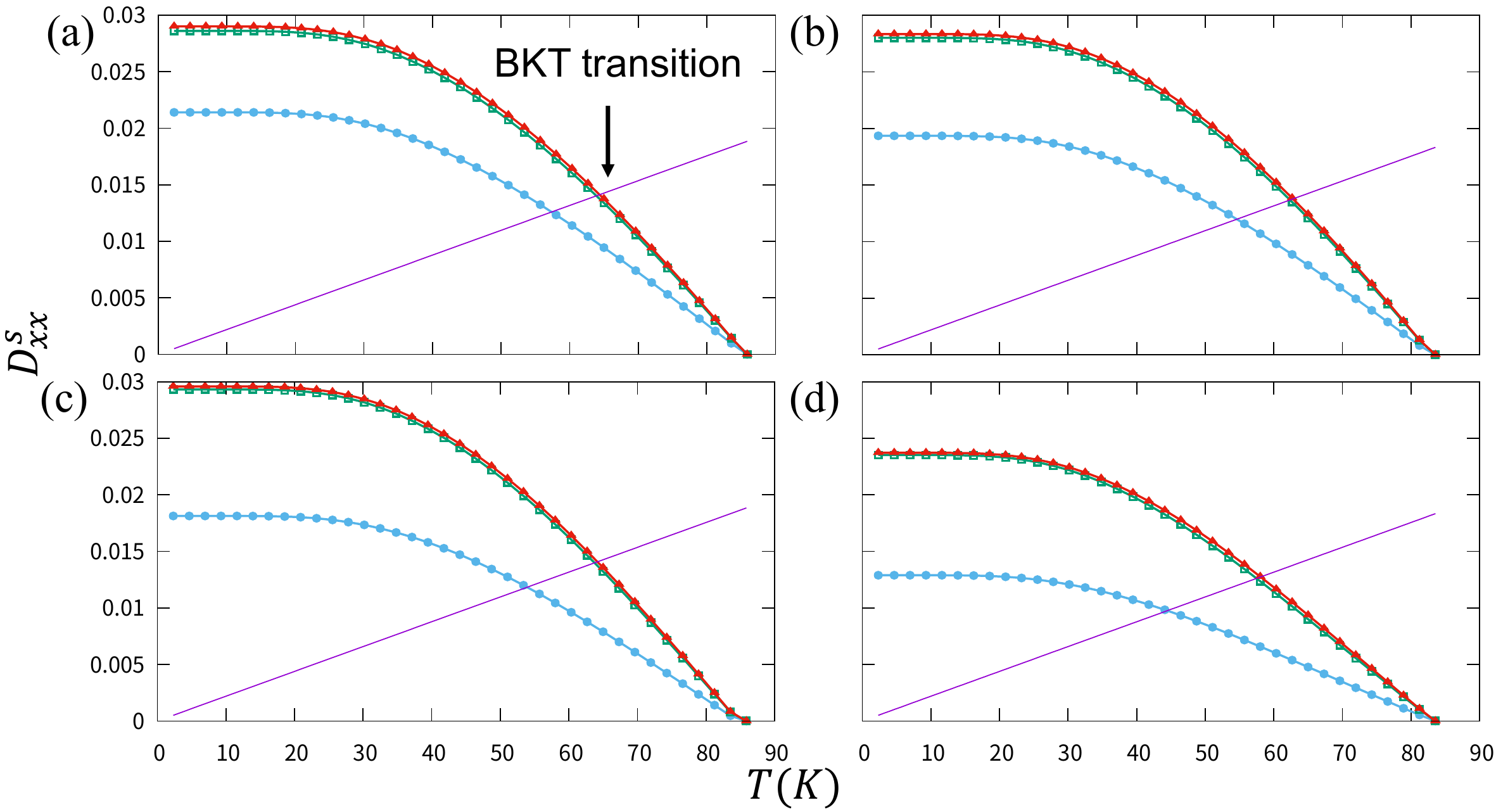}
  \centering
  \caption{Temperature dependence of the superfluid weight for $\bm k$-independent gap functions.
  The blue, green, and red lines show the conventional term ($D^{\rm conv}$), conventional + geometric term ($D^{\rm conv}+D^{\rm geom}$), and the total superfluid weight ($D^{\rm s}$), respectively.
  The red and green lines almost coincide because $D^{\rm multi}$ is negligible.
  The purple straight line shows $8T/\pi$, and the
  intersection with the red line determines the BKT transition temperature.
  We adopt the renormalization factor $z=1/5$ in (a) $n=6.1$, (b) $n=6.08$, and (c) $n=6.06$, 
  while $z = 1/8$ and $n = 6.06$ in (d).
  \label{fig:sfw_fese_83k}}
\end{figure}

\begin{figure}
    \centering
    \includegraphics[width=1.0\linewidth]{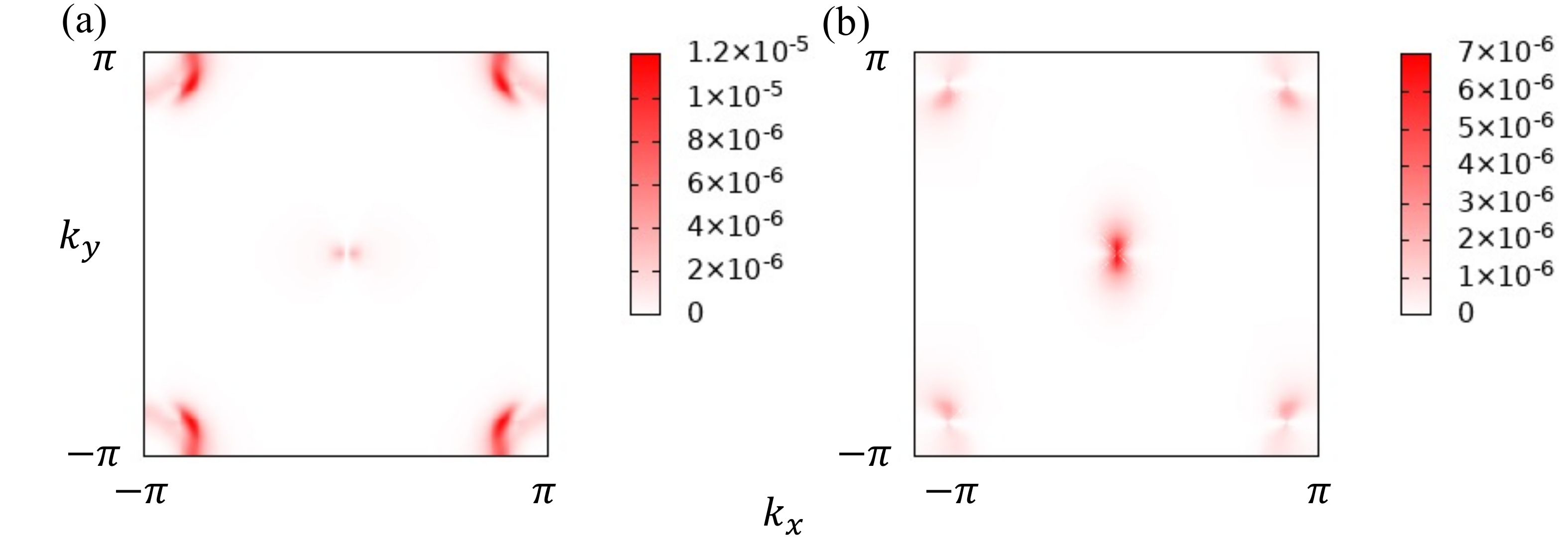}
    \caption{$\bm k$-resolved contribution to the superfluid weight from (a) the conventional term and (b) the geometric term. 
    The parameters are $n = 6.08$, $z = 1/5$, and $T \approx 2.3$~K. The $\bm k$-independent gap function is assumed.}
    \label{fig:sfw_kdep_n6.08}
\end{figure}

First, we show the results for the $\bm k$-independent gap function, corresponding to the isotropic $s$-wave superconductivity.
We consider an orbital-independent on-site pairing interaction, 
$
  V_{ij\bm k\bm k^\prime} = V_0\delta_{ij}
$.
The gap term $D^{\mathrm{gap}}_{\mu\nu}$ disappears in this case.
The temperature dependence of the superfluid weight 
is shown in Fig.~\ref{fig:sfw_fese_83k}.
Owing to the fourfold-rotational and mirror symmetries of the system, $D^{s}_{xx} = D^{s}_{yy}$ and $D^s_{xy}=0$ are satisfied.
Thus, the BKT transition temperature $T_{\rm BKT}$ is given by the relation $D^s_{xx}(T_{\rm BKT}) = 8T_{\rm BKT}/\pi$,
and the intersection between the purple straight line and the red solid line indicates the BKT transition temperature.
This is also valid for the $\bm{k}$-dependent gap functions discussed below.

For all parameter sets in Fig.~\ref{fig:sfw_fese_83k}, we see a significant geometric contribution 
to the superfluid weight, whereas the multi-gap term is negligible in all the results of this paper.
While the conventional term is suppressed as expected, the geometric term is enhanced by decreasing the electron number. This contrasting behavior leads to a particularly sizable contribution from the geometric term to the superfluid weight in the low electron-doping region. 
Accordingly, the geometric term enhances the BKT transition temperature by approximately $24\%$ 
for the case of $n = 6.06$ and $z = 1/8$, as shown in Fig.~\ref{fig:sfw_fese_83k}~(d).
The geometric term is furthermore significant at low temperatures because the geometric term is higher order in terms of $\Delta/E_{\rm F}$, where $\Delta$ is the magnitude of superconducting gap and $E_{\rm F}$ is the Fermi energy. 
We see that the geometric term determines nearly $45\%$ of the superfluid weight in Fig.~\ref{fig:sfw_fese_83k}~(d).
In this parameter set, the superconducting gap on the Fermi surface is approximately $10$~meV, which is consistent with 
the ARPES studies reporting the gap value from $8$~meV to $20$~meV~\cite{miyata2015high-temperature,wang2012interface-induced,xu2020spectroscopic,he2013phase}.
These results reveal that the superfluid weight in realistic monolayer FeSe is not determined sorely by the conventional term, and the geometric properties of Bloch wave functions play an essential role.

To obtain further insights, we show the ${\bm k}$-resolved contributions of the conventional and geometric terms in Fig.~\ref{fig:sfw_kdep_n6.08}. 
We see significant contributions from near the $M$ point in both terms, as expected from the presence of the Fermi surfaces.
Interestingly, there are also sizable contributions from near the $\Gamma$ point, where the Fermi surface is absent, and it is dominant in the geometric term.
This implies that the hole bands below the Fermi level have geometrically nontrivial properties and are essential for enhancing the superconductivity in monolayer FeSe. 

Next, we discuss the superconducting states of $\bm k$-dependent paring.
Here, we consider the pairing on the nearest- and next-nearest-neighbor bonds in addition to the on-site pairing.
The attractive interaction is assumed as
$
V_{ij\bm k\bm k^\prime} = V_0\delta_{ij}
+ V_1(\delta_{i,j+5} +\delta_{i+5,j})\cos k_x/2\cos k_y/2\cos k_x^\prime/2\cos k_y^\prime/2
+ V_2\delta_{ij}(\cos k_x +\cos k_y)(\cos k_x^\prime +\cos k_y^\prime),
$
where $V_1$ and $V_2$ represent the inter- and intra-sublattice attractive interactions, respectively.
The superconducting state belongs to the totally symmetric $A_{1g}$ representation irrespective of the parameters $V_0$, $V_1$, and $V_2$.
%
\begin{figure}[tbp]
  \includegraphics[width=1.0\linewidth]{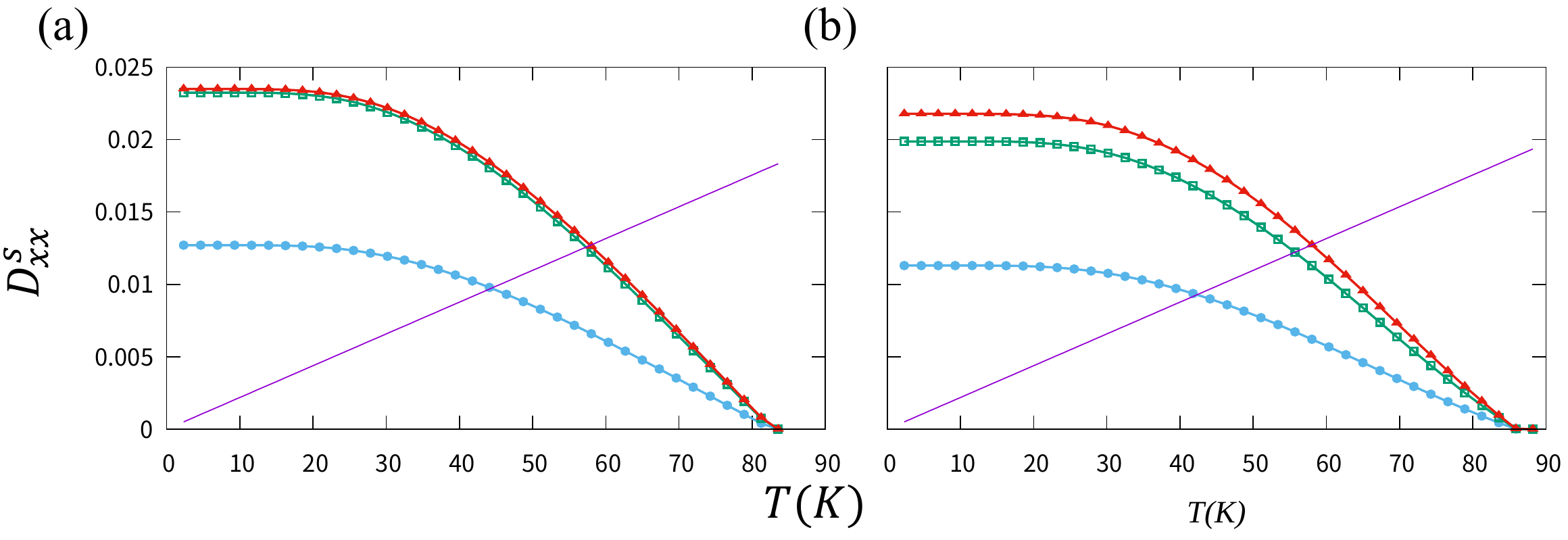}
  \centering
  \caption{
  Superfluid weight for the $\bm k$-dependent gap functions. 
  We set $z = 1/8$ and $n = 6.06$. The attractive interactions are (a) $V_1 = V_2 = 0.2V_0$ and (b) $V_1 = V_2 = 10V_0$. 
  The lines with different colors indicate the same quantities as Fig.~\ref{fig:sfw_fese_83k}.
}
\label{fig:sfw_fese_83_kdep_all}
\end{figure}
%
%
Figure~\ref{fig:sfw_fese_83_kdep_all} shows 
the superfluid weight for 
(a) $V_1 = V_2 = 0.2V_0$ and (b) $V_1 = V_2 = 10V_0$. Although the ${\bm k}$ dependence of gap function is not significant in Fig.~\ref{fig:sfw_fese_83_kdep_all}(a), Fig.~\ref{fig:sfw_fese_83_kdep_all}(b) shows the case of a highly $\bm k$-dependent gap function. 
Thus, Figs.~\ref{fig:sfw_fese_83_kdep_all}(a) and ~\ref{fig:sfw_fese_83_kdep_all}(b) correspond to the $s_{++}$-wave and incipient $s_{\pm}$-wave pairing states, respectively (see Supplemental Materials~\cite{Supplemental} for details).

In all figures, we can see significant geometric contributions to the superfluid weight
as we see in the case of the $\bm k$-independent pairing.
Therefore, we conclude that the geometric term plays an essential role in the superconductivity of monolayer FeSe regardless of pairing symmetry. 
On the other hand, the gap term shows a contrasting behavior between the $s_{++}$-wave pairing and the incipient $s_{\pm}$-wave pairing.
The gap term gives a non-negligible correction in the incipient $s_{\pm}$-wave state, although it is negligible in the $s_{++}$ state. 
The contribution to the superfluid weight by the gap term is about $8 \%$ and can not be ignored in the low-temperature regime of Fig.~\ref{fig:sfw_fese_83_kdep_all}~(b).
Thus, the $\bm k$-dependence in the gap function enhances the superfluid weight and BKT transition temperature through the gap term, which is different from the known effects of thermal excitation due to the anisotropic superconducting gap.

\textit{Conclusion.} ---
We formulated the superfluid weight of unconventional superconductors by taking into account the geometric term due to the nontrivial structure of Bloch wave functions and the gap term arising from the $\bm k$-dependence of the gap function.
Then, applying the formula to the model obtained by the first-principles calculation, we calculated the superfluid weight in monolayer FeSe. 
Via the superfluid weight, the geometric properties of Bloch electrons significantly enhance the superconductivity.
In particular, the geometric term enhances the BKT transition temperature by nearly $14$ K for a typical parameter set.
Furthermore, the $\bm k$-dependence of the gap function 
also stabilizes superconductivity in the incipient $s_{\pm}$-wave pairing state.

A surprisingly high transition temperature in monolayer FeSe is realized by two origins. One is the enhancement of mean-field transition temperature, probably owing to the excess electron doping or increased electron-phonon coupling. The other is the enhancement of BKT transition temperature attributed not only to the conventional Fermi liquid properties but also to the geometrically nontrivial properties arising from the multi-band structure. 
Unlike the previous studies on artificial systems with flat bands, FeSe does not contain a flat band. Instead of the flat band structure, the intriguing nature of superconductivity close to the BCS-BEC crossover makes the geometric properties play an essential role in FeSe.
The small Fermi surfaces reduce the conventional contribution, and interestingly, the nontrivial band dispersion near the $\Gamma$ point gives rise to a significant geometric contribution although the bands are below the Fermi level.
Thus, the monolayer FeSe is an intriguing platform in which the geometric properties of Bloch electrons enhance the superconductivity due to its intrinsic band structure.

\begin{acknowledgments}
We are grateful to K. Kimura, R. Ikeda and R. Sano for fruitful discussions and comments.
This work was supported by JSPS KAKENHI (Grants No.~JP18H05227, No.~JP18H01178, No.~20H05159, and No.~21K13880) and by 
SPIRITS 2020 of Kyoto University.
\end{acknowledgments}

\bibliography{main}

\end{document}


\newcommand{\rr}{{\bm r}}
\newcommand{\q}{{\bm q}}
\renewcommand{\k}{{\bm k}}
\newcommand*\wien    {\textsc{wien}2k\xspace}
\newcommand{\TK}[1]{{\color{red}{#1}}}

\newcommand*\YY[1]{\textcolor{blue}{#1}}
\newcommand{\YYS}[1]{\textcolor{blue}{\sout{#1}}}

\begin{center}
 {\large \textmd{Supplemental Materials:} \\[0.3em]
 {\bfseries Superconductivity in monolayer FeSe enhanced by quantum geometry}}
\end{center}

\begin{center}
\section{Derivation of the superfluid weight}
\end{center}

We derive the superfluid weight for unconventional superconductors, including $\bm k$-dependent Cooper pairs, using the Kubo formula and the Bardeen-Cooper-Schrieffer (BCS) mean-field theory.
For simplicity, we consider the time-reversal symmetric superconductors.
First, we derive a formula of the superfluid weight.
Then, we divide it into four terms.
Finally, we show that the conventional term added to the gap term is reduced to the known formula $n^*/m^*$ when
$\bm \Delta_{\bm k}^\dagger = \bm \Delta_{\bm k} = \bm 1 \times \Delta_{\bm k}$, while the geometric term is attributed to the band-resolved quantum metric.

\begin{center}
\end{center}
\begin{center}
\subsection{Superfluid weight via Kubo formula}
\end{center}

We start from a model with an attractive interaction,
\begin{eqnarray}
    \hat{H} = \sum_{\bm k}\sum_{ij\sigma}\hat{c}_{i\sigma\bm k}^\dagger h_{ij\bm k}\hat{c}_{j\sigma\bm k}
    + \sum_{\bm k \bm k^\prime}\sum_{ij}\hat{c}^\dagger_{i\uparrow\bm k}\hat{c}^\dagger_{j\downarrow-\bm k}V_{ij\bm k\bm k^\prime}\hat{c}_{j\downarrow-\bm k^\prime}\hat{c}_{i\uparrow\bm k^\prime},
\end{eqnarray}
where $h_{ij\bm k}$ is the Fourier transform of the hopping integral and
$V_{ij\bm k\bm k^\prime}$
represents an attractive interaction.
$\hat{c}_{i\sigma\bm k}^\dagger$ $(\hat{c}_{i\sigma\bm k})$
is the creation (annihilation) operator, and $i$, $\sigma$, and $\bm k$ represent the orbital and sublattice index, the spin index, and the wave vector, respectively.
We apply the BCS mean-field theory to the Hamiltonian,
\begin{eqnarray}
    \hat{H} &=& \sum_{\bm k}\sum_{ij\sigma}\hat{c}_{i\sigma\bm k}^\dagger h_{ij\bm k}\hat{c}_{j\sigma\bm k}
    + \sum_{\bm k}\sum_{ij}(\Delta_{ij\bm k}\hat{c}^\dagger_{i\uparrow\bm k}\hat{c}^\dagger_{j\downarrow-\bm k} + c.c.) - \sum_{\bm k\bm k^\prime}V_{ij\bm k\bm k^\prime}\braket{\hat{c}^\dagger_{i\uparrow\bm k}\hat{c}^\dagger_{j\downarrow-\bm k}}\braket{\hat{c}_{j\downarrow-\bm k^\prime}\hat{c}_{i\uparrow\bm k^\prime}}.\notag\\
\end{eqnarray}
Here, the gap function is determined by solving the gap equation,
\begin{eqnarray}
    \Delta_{ij\bm k} &=& \sum_{\bm k^\prime}V_{ij\bm k\bm k^\prime}\braket{\hat{c}_{j\downarrow-\bm k^\prime}\hat{c}_{i\uparrow\bm k^\prime}}.
\end{eqnarray}
For the superconductivity, we introduce the Nambu spinor,
\begin{eqnarray}
	\hat{\bm \psi}^{\dagger}_{\bm k} = \left(
	\begin{array}{cc}
		\hat{\bm c}^\dagger_{\uparrow\bm k},&
		\hat{\bm c}_{\downarrow-\bm k}
	\end{array}
	\right),\\
	\hat{\bm c}^{\dagger}_{\sigma\bm k} = \left(
	\begin{array}{ccc}
		\hat{c}^\dagger_{1\sigma\bm k},&\cdots,&
		\hat{c}^\dagger_{f\sigma\bm k}
	\end{array}
	\right),
\end{eqnarray}
and the Bogoliubov-de Gennes (BdG) Hamiltonian,
\begin{eqnarray}
    H_{{\rm BdG}\bm k} =
    \left(
	\begin{array}{cc}
		H_{0\bm k} & \bm \Delta_{\bm k}\\
		\bm \Delta^\dagger_{\bm k} & -H_{0-\bm k}^{T}
	\end{array}
    \right),\\
    H_{0\bm k} = \left(
    \begin{array}{cccc}
         h_{11\bm k}& h_{12\bm k} & \cdots &h_{1f\bm k}\\
         h_{21\bm k}& h_{22\bm k} & \cdots &h_{2f\bm k}\\
         \vdots& \vdots  & \ddots &\vdots\\
         h_{f1\bm k}& h_{f2\bm k} & \cdots &h_{ff\bm k}\\
    \end{array}
    \right).
\end{eqnarray}
Here, the dimension $f$ is the total number of the orbital and sublattice degrees of freedom.
In the time-reversal symmetric case, $H_{0\bm k} = H_{0-\bm k}^T$ is satisfied.
We can rewrite the mean-field Hamiltonian as,
\begin{eqnarray}
    \hat{H} = \sum_{\bm k} \hat{\bm \psi}_{\bm k}^\dagger H_{{\rm BdG}\bm k}\hat{\bm \psi}_{\bm k},
\end{eqnarray}
by ignoring the constant term.
We define the Nambu Green function as,
\begin{eqnarray}
	\mathcal{\bm G}(\bm k,\tau-\tau^\prime) &=&\braket{T_{\tau}[\hat{\bm \psi}^\dagger_{\bm k}(\tau)\otimes\hat{\bm \psi}_{\bm k}(\tau^\prime)]}.
\end{eqnarray}
Here, $\otimes$ represents the tensor product; the matrix elements for the normal part $i, j \leq f$ are defined by
$
		\mathcal{\bm G}_{ij}(\bm k,\tau-\tau^\prime) =
				\braket{T_{\tau}[\hat{c}^\dagger_{j\bm k\uparrow}(\tau)\hat{c}_{i\bm k\uparrow}(\tau^\prime)]}
$.
$T_{\tau}$ represents the time ordering product for the imaginary time $\tau$.

The superfluid weight, $D^{\rm s}_{\mu\nu}$, is defined as the $\bm q$-limit of the Meissner Kernel, $K_{\mu\nu}(\bm q,\omega)$,
\begin{eqnarray}
D_{\mu\nu}^{\rm  s}=\lim_{\bm q\rightarrow0}K_{\mu\nu}(\bm q,0).\label{eq:meisner_sfw}
\end{eqnarray}
First, we derive the Meissner Kernel, which represents the current response to the vector potential,
\begin{eqnarray}
	j_{\mu}(\bm q, \omega) = -K_{\mu\nu}(\bm q, \omega) A_\nu(\bm q, \omega).
\end{eqnarray}
Here, $j_{\mu}$ and $A_{\nu}$ are the current density and the vector potential, respectively.
In the linear response theory, the Meisner Kernel is obtained as,
\begin{eqnarray}
	&&K_{\mu\nu}(\bm q, \tau-\tau^\prime) =-K_{\mu\nu}^{\rm para}(\bm q, \tau-\tau^\prime) + K_{\mu\nu}^{\rm dia}(\bm q, \tau-\tau^\prime),\label{meisner}\\
	&&K_{\mu\nu}^{\rm para}(\bm q, \tau-\tau^\prime) =
	\sum_{\bm k\bm k^\prime}\sum_{\sigma\sigma^\prime}
	\Braket{T_{\tau}\left[
	\hat{\bm c}_{\sigma\bm k}^\dagger(\tau)
	\partial_{\mu}H_{0\bm k+\bm q/2}\hat{\bm c}_{\sigma\bm k+\bm q}(\tau)
	\hat{\bm c}_{\sigma^\prime\bm k^\prime}^\dagger(\tau^\prime)
	\partial_{\mu}H_{0\bm k^\prime-\bm q/2}\hat{\bm c}_{\sigma\bm k^\prime-\bm q}(\tau^\prime)
	\right]},\label{meisner_para}
	\notag\\&&
	\\
	&&K_{\mu\nu}^{\rm dia}(\bm q, \tau-\tau^\prime) =
	\sum_{\bm k}\sum_{\sigma}
	\Braket{T_{\tau}\left[\hat{\bm c}_{\sigma\bm k}^\dagger(\tau)
	\partial_{\mu}\partial_{\nu}H_{0\bm k+\bm q/2}\hat{\bm c}_{\sigma\bm k+\bm q}(\tau)\right]}\delta(\tau- \tau^\prime)\label{meisner_dia}.
\end{eqnarray}
$K^{\rm para}_{\mu\nu}$ is the paramagnetic term, and
$K^{\rm dia}_{\mu\nu}$ is the diamagnetic term.
Using the block diagonal Hamiltonian,
\begin{eqnarray}
    H_{p(m)\bm k} = \left(
    \begin{array}{cc}
         H_{0\bm k}& 0\\
         0 & (-)H_{0-\bm k}^{T}\\
    \end{array}
    \right),
\end{eqnarray}
these terms are written by,
\begin{eqnarray}
 	&&K_{\mu\nu}^{\rm para}(\bm q, \tau-\tau^\prime) =\sum_{\bm k\bm k^\prime}
	\Braket{T_{\tau}\left[
	\hat{\bm \psi}_{\bm k}^\dagger(\tau)
	\partial_{\mu}H_{p\bm k+\bm q/2}\hat{\bm \psi}_{\bm k+\bm q}(\tau)
	\hat{\bm \psi}_{\bm k^\prime}^\dagger(\tau^\prime)
	\partial_{\mu}H_{p\bm k^\prime-\bm q/2}\hat{\bm \psi}_{\bm k^\prime-\bm q}(\tau^\prime)
	\right]},\label{meisner_para2}\\
	&&K_{\mu\nu}^{\rm dia}(\bm q, \tau-\tau^\prime) =
	\sum_{\bm k}
	\Braket{T_{\tau}\left[\hat{\bm \psi}_{\bm k}^\dagger(\tau)
	\partial_{\mu}\partial_{\nu}H_{m\bm k+\bm q/2}\hat{\bm \psi}_{\bm k+\bm q}(\tau)\right]}\delta(\tau- \tau^\prime)\label{meisner_dia2}.
\end{eqnarray}
In the Gorkov approximation in which we ignore the vertex correction, these terms are given by the Green function,
\begin{eqnarray}
    &&K_{\mu\nu}^{\rm para}(\bm q,\tau-\tau^\prime) = -\sum_{\bm k}{\rm Tr}\left[
	\partial_{\mu}H_{p\bm k+\bm q/2}\mathcal{\bm G}(\bm k,\tau-\tau^\prime)
	\partial_{\nu}H_{p\bm k+\bm q/2}\mathcal{\bm G}(\bm k+\bm q,\tau^\prime-\tau)
	\right],\\
	&&K_{\mu\nu}^{\rm dia}(\bm q, \tau- \tau^\prime) = \sum_{\bm k}{\rm Tr}\left[
	\partial_{\mu}\partial_{\nu}H_{m\bm k+\bm q/2}\mathcal{\bm G}(\bm k ,0)\delta(\tau- \tau^\prime)
	\right].
\end{eqnarray}
Trace is taken for the orbital, sublattice, and particle-hole degrees of freedom.
After the Fourier transform with respect to the imaginary time, we have
\begin{eqnarray}
    K_{\mu\nu}^{\rm para}(\bm q,\Omega) &=& -\sum_{\bm k, \,\omega}{\rm Tr}\left[
	\partial_{\mu}H_{p\bm k+\bm q/2}\mathcal{\bm G}(\bm k,\omega)
	\partial_{\nu}H_{p\bm k+\bm q/2}\mathcal{\bm G}(\bm k+\bm q,\omega-\Omega)
	\right], \\
	K_{\mu\nu}^{\rm dia}(\bm q, \Omega) &=& \sum_{\bm k, \,\omega}{\rm Tr}\left[
	\partial_{\mu}\partial_{\nu}H_{m\bm k+\bm q/2}\mathcal{\bm G}(\bm k ,\omega)
	\right],
\end{eqnarray}
where $\omega$ and $\Omega$ are the Fermionic and Bosonic Matsubara frequencies, respectively.
The Nambu Green function with frequency $\omega$ is written as,
\begin{eqnarray}
    \mathcal{G}(\bm k,\omega) = \dfrac{1}{i\omega-H_{{\rm BdG}\bm k}}.
\end{eqnarray}
Using the eigenvalue equation for the BdG Hamiltonian, $H_{{\rm BdG}\bm k}\ket{\psi_{\alpha\bm k}} = E_{\alpha\bm k}\ket{\psi_{\alpha\bm k}}$, and taking the limit $\Omega = 0, \bm q\rightarrow0$, we get the formula of the superfluid weight,
\begin{eqnarray}
    D_{\mu\nu}^{\rm s} &=& -D_{\mu\nu}^{\rm para} + D_{\mu\nu}^{\rm dia},\\
    D_{\mu\nu}^{\rm para} &=& -\sum_{\bm k}\sum_{\alpha\beta}\dfrac{f(E_{\alpha\bm k})-f(E_{\beta\bm k})}{E_{\alpha\bm k}-E_{\beta\bm k}}
	\bra{\psi_{\alpha\bm k}}\partial_{\mu}H_{p\bm k}\ket{\psi_{\beta\bm k}}\bra{\psi_{\beta\bm k}}\partial_{\mu}H_{p\bm k}\ket{\psi_{\alpha\bm k}}
	,\label{eq:Ds_para_1}\\
	D_{\mu\nu}^{\rm dia} &=& \sum_{\bm k}\sum_{\alpha}f(E_{\alpha\bm k})
	\bra{\psi_{\alpha\bm k}}\partial_{\mu}\partial_{\nu}H_{m\bm k}\ket{\psi_{\alpha\bm k}}.\label{eq:Ds_dia_1}
\end{eqnarray}
Equation~\eqref{eq:Ds_dia_1} for the diamagnetic term can be rewritten as,
\begin{align}
   	D_{\mu\nu}^{\rm dia}
	=&-\sum_{\bm k}\sum_{\alpha}\left(\partial_{\mu}f(E_{\alpha\bm k})
	\bra{\psi_{\alpha\bm k}}\partial_{\nu}H_{m\bm k}\ket{\psi_{\alpha\bm k}}
	\right.\notag\\
	&+f(E_{\alpha\bm k})
	\bra{\partial_{\mu}\psi_{\alpha\bm k}}\partial_{\nu}H_{m\bm k}
	\ket{\psi_{\alpha\bm k}}
	+f(E_{\alpha\bm k})
	\left.
	\bra{\psi_{\alpha\bm k}}\partial_{\nu}H_{m\bm k}
	\ket{\partial_{\mu}\psi_{\alpha\bm k}}
	\right), \notag\\
	=&-\sum_{\bm k}\sum_{\alpha}\left(\partial_{\mu}f(E_{\alpha\bm k})
	\bra{\psi_{\alpha\bm k}}\partial_{\nu}H_{m\bm k}\ket{\psi_{\alpha\bm k}}
	\right.\notag\\
	&+f(E_{\alpha\bm k})
	\sum_{\beta(\neq \alpha)}\braket{\partial_{\mu}\psi_{\alpha\bm k}\vert\psi_{\beta\bm k}}\bra{\psi_{\beta\bm k}}\partial_{\nu}H_{m\bm k}
	\ket{\psi_{\alpha\bm k}}
	+f(E_{\alpha\bm k})
	\sum_{\beta(\neq \alpha)}\bra{\psi_{\alpha\bm k}}\partial_{\nu}H_{m\bm k}
	\ket{\psi_{\beta\bm k}}\braket{\psi_{\beta\bm k}\vert\partial_{\mu}\psi_{\alpha\bm k}}
	), \notag\\
	=&-\sum_{\bm k}\sum_{\alpha\beta}\dfrac{f(E_{\alpha\bm k})-f(E_{\beta\bm k})}{E_{\alpha\bm k}-E_{\beta\bm k}}
	\bra{\psi_{\alpha\bm k}}\partial_{\nu}H_{m\bm k}\ket{\psi_{\beta\bm k}}\bra{\psi_{\beta\bm k}}\partial_{\mu}H_{{\rm BdG}\bm k}\ket{\psi_{\alpha\bm k}}, \notag\\
	=&-\sum_{\bm k}\sum_{\alpha\beta}\dfrac{f(E_{\alpha\bm k})-f(E_{\beta\bm k})}{E_{\alpha\bm k}-E_{\beta\bm k}}
	\bra{\psi_{\alpha\bm k}}\partial_{\nu}H_{m\bm k}\ket{\psi_{\beta\bm k}}\bra{\psi_{\beta\bm k}}\partial_{\mu}H_{m\bm k}\ket{\psi_{\alpha\bm k}}
	\notag\\
	&-\sum_{\bm k}\sum_{\alpha\beta}\dfrac{f(E_{\alpha\bm k})-f(E_{\beta\bm k})}{E_{\alpha\bm k}-E_{\beta\bm k}}
	\bra{\psi_{\alpha\bm k}}\partial_{\nu}H_{m\bm k}\ket{\psi_{\beta\bm k}}\bra{\psi_{\beta\bm k}}\partial_{\mu}
	\left(
	\begin{array}{cc}
	    0 & \bm \Delta_{\bm k} \\
	    \bm \Delta_{\bm k}^\dagger & 0
	\end{array}
	\right)
	\ket{\psi_{\alpha\bm k}}.\label{eq:Ds_dia_2}
\end{align}
Here, we use the relationship,
 $\braket{\partial_{\mu}\psi_{\alpha\bm k}\vert\psi_{\alpha\bm k}} +\braket{\psi_{\alpha\bm k}\vert\partial_{\mu}\psi_{\alpha\bm k}}= 0$ since $\braket{\psi_{\alpha\bm k}\vert\psi_{\alpha\bm k}} = 1$.

\begin{center}
\end{center}
\begin{center}
\subsection{Superfluid weight based on geometric properties of Bloch electrons}
\end{center}
Next, we divide the superfluid weight into four terms based on geometric properties of Bloch electrons.
First, we rewrite $\ket{\psi_{\alpha\bm k}}$ using the normal state Bloch wave function.
For this purpose, we use the eigenvalue equation of the normal Hamiltonian $H_{0\bm k}\ket{u_{n\bm k}} = \epsilon_{n\bm k}\ket{u_{n\bm k}}$.
The unitary matrix which can diagonalize the normal Hamiltonian is written as,
\begin{eqnarray}
    U_{\bm k} = \left(
    \begin{array}{cccc}
        \ket{u_{1\bm k}}, & \ket{u_{2\bm k}}, & \cdots, &\ket{u_{f\bm k}}\\
    \end{array}
    \right).
\end{eqnarray}
Using this unitary matrix, we get the BdG Hamiltonian of the band representation,
\begin{eqnarray}
    \left(
	\begin{array}{cc}
		U_{\bm k}^{\dagger} & 0\\
		0 & U_{\bm k}^{\dagger}
	\end{array}
    \right)
    H_{{\rm BdG}\bm k}
    \left(
	\begin{array}{cc}
		U_{\bm k} & 0\\
		0 & U_{\bm k}
	\end{array}
    \right)
    =
    \left(
	\begin{array}{cc}
		U_{\bm k}^{\dagger}H_{0\bm k}U_{\bm k} & U_{\bm k}^{\dagger}\bm\Delta_{\bm k}U_{\bm k}\\
		U_{\bm k}^{\dagger}\bm \Delta^\dagger_{\bm k}U_{\bm k} & -U_{\bm k}^{\dagger}H_{0\bm k}U_{\bm k}
	\end{array}
    \right).\label{eq:u_h_bdg_u}
\end{eqnarray}
We consider the unitary matrix which diagonalizes Eq.~\eqref{eq:u_h_bdg_u},
\begin{eqnarray}
	\Phi_{\bm k} = \left(
	\begin{array}{cccc}
		\phi_{1\bm k}^{1\uparrow}& \phi_{1\bm k}^{2\uparrow} & \cdots & \phi_{1\bm k}^{M\uparrow} \\
		\phi_{2\bm k}^{1\uparrow} & \phi_{2\bm k}^{2\uparrow} & \cdots & \phi_{2\bm k}^{M\uparrow} \\
		\vdots & \vdots & \ddots & \vdots \\
		\phi_{f\bm k}^{1\uparrow} & \phi_{f\bm k}^{2\uparrow} & \cdots & \phi_{f\bm k}^{M\uparrow} \\
		\phi_{1\bm k}^{1\downarrow} & \phi_{1\bm k}^{2\downarrow} & \cdots & \phi_{1\bm k}^{M\downarrow} \\
		\phi_{2\bm k}^{1\downarrow} & \phi_{2\bm k}^{2\downarrow} & \cdots & \phi_{2\bm k}^{M\downarrow} \\
		\vdots & \vdots & \ddots & \vdots \\
		\phi_{f\bm k}^{1\downarrow} & \phi_{f\bm k}^{2\downarrow} & \cdots & \phi_{f\bm k}^{M\downarrow}
	\end{array}
	\right),
\end{eqnarray}
with $M = 2\times f$.
Accordingly, we represent $\ket{\psi_{\alpha\bm k}}$ with $\ket{u_{n\bm k}}$,
\begin{eqnarray}
   	\ket{\psi_{\alpha\bm k}} =\left(
	\begin{array}{c}
		\sum_n\phi_{n\bm k}^{\alpha\uparrow}\ket{u_{n\bm k}}\\
		\sum_n\phi_{n\bm k}^{\alpha\downarrow}\ket{u_{n\bm k}}
	\end{array}
	\right).\label{eq:psi_u}
\end{eqnarray}
Inserting Eq.~\eqref{eq:psi_u} to Eq.~\eqref{eq:Ds_para_1} and the first term of Eq.~\eqref{eq:Ds_dia_2}, we get
\begin{eqnarray}
   &&2\sum_{\bm k}\sum_{\alpha\beta}\sum_{nmls}\dfrac{f(E_{\alpha\bm k})-f(E_{\beta\bm k})}{E_{\alpha\bm k}-E_{\beta\bm k}}
  \phi_{n\bm k}^{*\alpha\uparrow}\phi_{m\bm k}^{\beta\uparrow}\phi_{l\bm k}^{*\beta\downarrow}\phi_{s\bm k}^{\alpha\downarrow}
  (\bra{u_{n\bm k}}\partial_{\mu}H_{0\bm k}\ket{u_{m\bm k}}\bra{u_{l\bm k}}\partial_{\nu}H_{0\bm k}\ket{u_{s\bm k}}
	+(\mu\leftrightarrow\nu)).\notag\\
\end{eqnarray}
From the perspective of the interband and intraband contributions, this is divided into three terms,
\begin{eqnarray}
    &&D^{\rm conv}_{\mu\nu} = 2\sum_{nm\bm k}C_{nnmm\bm k}^{\uparrow\uparrow\downarrow\downarrow}(\partial_\mu\epsilon_{n\bm k}\partial_\nu\epsilon_{m\bm k}+(\mu\leftrightarrow\nu)),\label{eq:conv}\\
    &&D^{\rm geom}_{\mu\nu} = 2\sum_{n\neq m, l\neq s, {\bm k}}C_{nmls\bm k}^{\uparrow\uparrow\downarrow\downarrow}(\epsilon_{n\bm k}-\epsilon_{m\bm k})(\epsilon_{l\bm k}-\epsilon_{s\bm k})(\braket{\partial_{\mu}u_{n\bm k}\vert u_{m\bm k}}\braket{u_{l\bm k}\vert\partial_{\nu}u_{s\bm k}} + (\mu\leftrightarrow\nu)),\label{eq:geom}\\
    &&D^{\rm multi}_{\mu\nu} = 2\sum_{n, l\neq s, {\bm k}}\left(C_{nnls\bm k}^{\uparrow\uparrow\downarrow\downarrow}\partial_\mu\epsilon_{n\bm k}(\epsilon_{l\bm k}-\epsilon_{s\bm k})\braket{\partial_{\nu}u_{l\bm k}\vert u_{s\bm k}}
    +C_{lsnn\bm k}^{\uparrow\uparrow\downarrow\downarrow}\partial_\nu\epsilon_{n\bm k}(\epsilon_{l\bm k}-\epsilon_{s\bm k})\braket{\partial_{\mu}u_{l\bm k}\vert u_{s\bm k}}+(\mu\leftrightarrow\nu)\right).\notag\\\label{eq:aniso}
\end{eqnarray}
Here, we define
\begin{eqnarray}
    C_{nmls\bm k}^{\sigma_1\sigma_2\sigma_3\sigma_4} = \sum_{\alpha\beta\bm k}\dfrac{f(E_{\alpha\bm k})-f(E_{\beta\bm k})}{E_{\alpha\bm k}-E_{\beta\bm k}}
  \phi_{n\bm k}^{\alpha\sigma_1*}\phi_{m\bm k}^{\beta\sigma_2}\phi_{l\bm k}^{\beta\sigma_3*}\phi_{s\bm k}^{\alpha\sigma_4}.
\end{eqnarray}

Equation~\eqref{eq:conv} shows the conventional term $D^{\rm conv}_{\mu\nu}$, which arises from the intraband effect.
On the other hand, Eq.~\eqref{eq:geom} denotes the geometric term $D^{\rm geom}_{\mu\nu}$, which is purely the interband effect.
It should be noticed that $D_{\mu\nu}^{\rm conv}$ is determined by the band dispersion, while $D_{\mu\nu}^{\rm geom}$ reflects the geometric properties of the Bloch wave functions.
In addition to these terms, we obtain the multi-gap term, $D_{\mu\nu}^{\rm multi}$ in Eq.~\eqref{eq:aniso}, which comes from the interband pairing effect.
In the absence of the interband pairing, this term vanishes.
In previous studies, this term has been included in the geometric term~\cite{julku2020superfluid}.
However, from the perspective of the interband and intraband effects, we distinguish this term from the geometric term.
Equation~\eqref{eq:aniso} for the multi-gap term reveals that both intraband and interband effects are needed.

Finally, inserting Eq.~\eqref{eq:psi_u} to the second term of Eq.~\eqref{eq:Ds_dia_2}, we get the gap term,
\begin{align}
    D^{\rm gap}_{\mu\nu} =
    \sum_{nmls\sigma {\bm k}}
    S\left(C_{nmls\bm k}^{\uparrow\downarrow\sigma\sigma}\bra{u_{n\bm k}}\partial_\mu \bm\Delta_{\bm k}\ket{u_{m\bm k}}+C_{nmls\bm k}^{\downarrow\uparrow\sigma\sigma}\bra{u_{n\bm k}}\partial_\mu \bm\Delta^\dagger_{\bm k}\ket{u_{m\bm k}}\right)
    \bra{u_{l\bm k}}\partial_{\nu}H_{0\bm k}\ket{u_{s\bm k}} \label{eqgap},
\end{align}
which comes from the $\bm k$-dependence of the gap function.
$S$ takes $-(+)$ when $\sigma = \uparrow(\downarrow)$.
Thus, we can divide the superfluid weight into the four tems as $D_{\mu\nu}^{s} = D_{\mu\nu}^{\rm conv}+ D_{\mu\nu}^{\rm geom} + D_{\mu\nu}^{\rm multi}+ D_{\mu\nu}^{\rm gap}$.

\begin{center}
\end{center}
\begin{center}
\subsection{Superfluid weight in the case of $\Delta_{k} = diag(\Delta_{k})$}
\end{center}

We can simplify the formula for $D_{\mu\nu}^{\rm conv} + D_{\mu\nu}^{\rm gap}$ and $D_{\mu\nu}^{\rm geom}$ when
$\bm \Delta_{\bm k} = \bm 1 \times \Delta_{\bm k}$. We take a real-valued $\Delta_{\bm k}$ without loss of generality.
In this case, the unitary matrix which diagonalizes the BdG Hamiltonian of the band representation can be written in a simple form,
\begin{eqnarray}
	\phi_{n\bm k}^{i\uparrow} = u_{n\bm k}\delta_{n,i}-v_{n\bm k}\delta_{n+f,i},\\
	\phi_{n\bm k}^{i\downarrow} = v_{n\bm k}\delta_{n,i} + u_{n\bm k}\delta_{n+f,i},
\end{eqnarray}
with
\begin{eqnarray}
	u_{n\bm k} = \dfrac{1}{\sqrt{2}}\sqrt{1 + \dfrac{\epsilon_{n\bm k}}{E_{n\bm k}}}, \hspace{5mm}
	v_{n\bm k} = \dfrac{1}{\sqrt{2}}\sqrt{1 - \dfrac{\epsilon_{n\bm k}}{E_{n\bm k}}},
\end{eqnarray}
and $E_{n\bm k} = \sqrt{\epsilon_{n\bm k}^2 + \Delta^2_{\bm k}}$.
In this case, the multi-gap term vanishes since $C_{nmls\bm k}^{\sigma_1\sigma_2\sigma_3\sigma_4}$
is finite only for $n = s, m = l$.
Using the above simplification, $D_{\mu\nu}^{\rm conv}$, $D_{\mu\nu}^{\rm geom}$, and $D_{\mu\nu}^{\rm gap}$ are written as,
\begin{align}
    D_{\mu\nu}^{\rm conv} &= - \sum_{n\sigma\bm k}\bra{u_{n\bm k}}\partial_{\mu}H_{0\bm k}\ket{u_{n\bm k}}\bra{u_{n\bm k}}\partial_{\nu}H_{0\bm k}\ket{u_{n\bm k}}
    \left(
    \sigma\dfrac{\vert\Delta_{\bm  k}\vert^2}{E_{n\bm k}^3}f(\sigma E_{n\bm k})-
    \dfrac{\vert\Delta_{\bm  k}\vert^2}{E_{n\bm k}^2}f^\prime(\sigma E_{n\bm k})
    \right),\label{eq:Ds_conv_ts}\\
    D_{\mu\nu}^{\rm geom} &= \sum_{n\neq m\sigma\sigma^\prime\bm k}\bra{u_{n\bm k}}\partial_{\mu}H_{0\bm k}\ket{u_{m\bm k}}\bra{u_{m\bm k}}\partial_{\nu}H_{0\bm k}\ket{u_{n\bm k}}
    \dfrac{f(\sigma E_{n\bm k})-f(\sigma^\prime E_{m\bm k})}{\sigma E_{n\bm k}-\sigma^\prime E_{m\bm k}}
    \left(
    \sigma\sigma^\prime\dfrac{\vert\Delta_{\bm  k}\vert^2}{E_{n\bm k}E_{m\bm k}}
    \right),\label{eq:Ds_geom_ts}\notag\\\\
    D_{\mu\nu}^{\rm gap} &= - \sum_{n\sigma\bm k}\bra{u_{n\bm k}}\partial_{\mu}\bm \Delta_{\bm k}\ket{u_{n\bm k}}\bra{u_{n\bm k}}\partial_{\nu}H_{0\bm k}\ket{u_{n\bm k}}
    \left(
    -\sigma\dfrac{\Delta_{\bm k}\epsilon_{n\bm k}}{E_{n\bm k}^3}f(\sigma E_{n\bm k})+
    \dfrac{\Delta_{\bm k}\epsilon_{n\bm k}}{E_{n\bm k}^2}f^\prime(\sigma E_{n\bm k})
    \right),\notag\\\label{eq:Ds_gap_ts}
\end{align}
where $\sigma$ takes $\pm1$.

At zero temperature, $D^{\rm conv}_{\mu\nu} + D^{\rm gap}_{\mu\nu}$ becomes
\begin{eqnarray}
    D_{\mu\nu}^{\rm conv} + D_{\mu\nu}^{\rm gap}&=& \sum_{n\bm k}\left(
    \partial_{\mu}\epsilon_{n\bm k}\partial_{\nu}\epsilon_{n\bm k}
    \dfrac{\vert\Delta_{\bm  k}\vert^2}{E_{n\bm k}^3}
    - \partial_{\mu}\Delta_{\bm k}\partial_{\nu}\epsilon_{n\bm k}
    \dfrac{\Delta_{\bm  k}\epsilon_{n\bm k}}{E_{n\bm k}^3}
    \right), \notag\\
    &=&\sum_{n\bm k}\partial_{\nu}\epsilon_{n\bm k}\partial_{\mu}
    \left(\dfrac{\epsilon_{n \bm k}}{E_{n\bm k}}\right), \notag\\
    &=&-\sum_{n\bm k}\partial_{\nu}\epsilon_{n\bm k}\partial_{\mu}
    \left(1-\dfrac{\epsilon_{n \bm k}}{E_{n\bm k}}\right),\notag\\
    &=&\sum_{n\bm k}\partial_{\nu}\partial_{\mu}\epsilon_{n\bm k}
    \left(1-\dfrac{\epsilon_{n \bm k}}{E_{n\bm k}}\right).\label{eq:m_over_n}
\end{eqnarray}
Note that $(1-\dfrac{\epsilon_{n \bm k}}{E_{n\bm k}})$ is the expected value of the particle number and $\partial_{\nu}\partial_{\mu}\epsilon_{n\bm k}$ is the inverse mass tensor.
Thus, $D_{\mu\nu}^{\rm conv} + D_{\mu\nu}^{\rm geom}$ reduces to $n^{*}/m^{*}$.
In addition, the geometric term $D_{\mu\nu}^{\rm geom}$ is written as,
\begin{eqnarray}
    D_{\mu\nu}^{\rm geom} &=& \dfrac{1}{2}\sum_{nm\sigma\sigma^\prime\bm k}(\epsilon_{n\bm k}-\epsilon_{m\bm k})^2\left(\braket{\partial_{\mu}u_{n\bm k}\vert u_{m\bm k}}\braket{u_{m\bm k}\vert\partial_{\nu}u_{n\bm k}}+ c.c.\right)\notag\\
    &\times&\dfrac{f(\sigma E_{n\bm k})-f(\sigma^\prime E_{m\bm k})}{\sigma E_{n\bm k}-\sigma^\prime E_{m\bm k}}
    \left(
    \sigma\sigma^\prime\dfrac{\vert\Delta_{\bm  k}\vert^2}{E_{n\bm k}E_{m\bm k}}
    \right),\label{eq:geom_metric}
\end{eqnarray}
where $\left(\braket{\partial_{\mu}u_{n\bm k}\vert u_{m\bm k}}\braket{u_{m\bm k}\vert\partial_{\nu}u_{n\bm k}}+ c.c.\right)$ is the band-resolved quantum metric.
In the isolated-band limit, the geometric term is reduced to the quantum metric in an original sense~\cite{liang2017band}.
We note that Eq. \eqref{eq:m_over_n} and Eq. \eqref{eq:geom_metric} are valid even for $\bm \Delta_{\bm k} \ne \bm 1 \times \Delta_{\bm k}$ when the gap function is $\bm k$-independent and the interband pairing is absent.
In this case, $E_{n\bm k} = \sqrt{\epsilon_{n\bm k}^2 + \vert\Delta_{n}\vert^2}$ and $\Delta_{n} = \bra{u_{n\bm k}}\bm \Delta\ket{u_{n\bm k}}$.

\begin{center}
\end{center}
\begin{center}
\section{Tight-binding model reproducing Fermi surfaces of FeSe}
\end{center}

The angle-resolved photoemission spectroscopy (ARPES) experiments observed two hole-like Fermi surfaces at the $\Gamma$ point and two electron-like Fermi surfaces at the $M$ point in bulk FeSe~\cite{maletz2014unusual,zhang2016distinctive}.
However, in the first-principles calculations, the Fermi surfaces are larger than those observed in the ARPES, and an extra Fermi surface is predicted. Thus, there is a slight discrepancy between the calculations and experiments.

To reproduce the Fermi surfaces of the bulk FeSe observed in the experiments, we slightly modify the hopping parameters given by the first-principles calculation~\cite{yamakawa2016nematicity,ohnari2016sign-reversing,ishizuka2018fermi}.
For this purpose, the energies of the $d_{xy}$-orbital band and the $d_{xz/yz}$-orbital band are shifted by ($-0.28,0,0.20$) and ($-0.27,0,0.13$) at the ($\Gamma$, $X$, $M$) points in the folded Brillouin zone, respectively.
To realize the energy shift, the hopping parameters are changed so as to satisfy
\begin{align}
    \delta E_{l} (\Gamma) &= \delta t_{ll}^{\rm on-site} +
    4\delta t_{ll}^{\rm nn} + 4\delta t_{ll}^{\rm nnn},\\
    \delta E_l (X) &= \delta t_{ll}^{\rm on-site},\\
    \delta E_l (M) &= \delta t_{ll}^{\rm on-site} -
    4\delta t_{ll}^{\rm nnn},
\end{align}
where we represent the energy shifts of the $l$-orbital band at the $\Gamma$, $X$ and $M$ points
as $\delta E_l(\Gamma), \delta E_l(X)$ and $\delta E_l(M)$, respectively. The modification in the intra-orbital hopping integral is represented by
$\delta t_{ll}$, and "on-sine", "nn", and "nnn" denote
the on-sine, first nearest-neighbour, and second nearest-neighbour hoppings, respectively.
In the 10-orbital model with two sublattices in the unit cell, $\delta t_{ll}^{\rm nn}$ ($\delta t_{ll}^{\rm nnn}$) is the inter-sublattice (intra-sublattice) hopping.
We also tune the chemical potential so that the extra Fermi surface near the $\Gamma$ point vanishes and the fillings are $n=6.06, 6.08, 6.1$.
Using these parameters, we can reproduce the Fermi surfaces of monolayer FeSe grown on SrTiO$_3$.

\begin{center}
\end{center}
\begin{center}
\section{Superfluid weight for $T_{\rm c} = 65K$ with the first-principles model}
\end{center}

\begin{figure}[htbp]
  \includegraphics[width=1.0\linewidth]{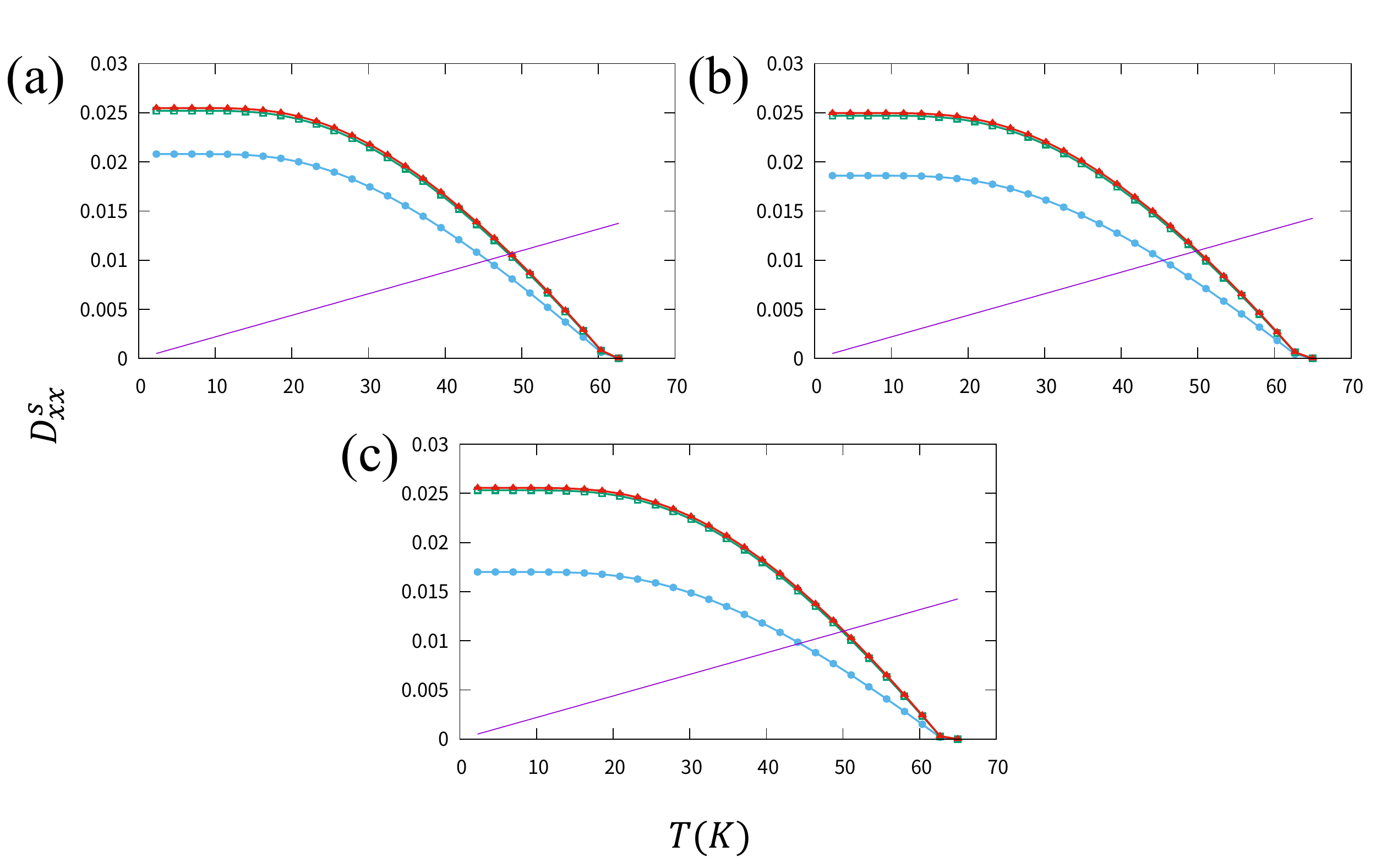}
  \centering
  \caption{Temperature dependence of the superfluid weight for $\bm k$-independent pairing and $z = 1/5$.
  The blue, green, and red lines show the conventional term ($D^{\rm conv}$), conventional + geometric term ($D^{\rm geom} +D^{\rm conv}$), and the total superfluid weight ($D^{\rm s}$), respectively.
  The purple line shows $8T/\pi$.
  The intersection between the purple line and the red line shows the BKT transition temperature.
  (a), (b), and (c) are for $n = 6.1$, $6.08$, and $6.06$, respectively.
  \label{fig:sfw_fese_65k_m5.0_all}}
\end{figure}

In this section, we show the superfluid weight for the mean-field superconducting transition temperature $T_{\rm c} \approx 65$~K using the 10-orbital tight-binding model.
In the main text we show the results for $T_{\rm c} \approx 83$~K. Below we see qualitatively the same results.

For simplicity, we consider the $\bm k$-independent pairing, given by an on-site pairing interaction $V_{ij\bm k} = V_0\delta_{ij}$, in this section.
The temperature dependence of the superfluid weight is shown in Fig.~\ref{fig:sfw_fese_65k_m5.0_all}.
In all panels, the geometric term gives a sizable correction to the superfluid weight.
In Fig.~\ref{fig:sfw_fese_65k_m5.0_all} (b), the geometric term determines the superfluid weight about $32\%$ at $T \approx 2.3$~K.
The magnitude of the gap function on the Fermi surface is nearly $9.3$~meV in Fig.~\ref{fig:sfw_fese_65k_m5.0_all} (b), while its experimental values are reported as at most $20$~meV~\cite{wang2012interface-induced} and at least $8$~meV~\cite{miyata2015high-temperature}.
Thus, a larger gap function may be realized in monolayer FeSe.
For a larger $T_{\rm c}$ and gap function, a more significant geometric contribution to the superfluid weight is obtained, as we see in the main text.

\begin{center}
\end{center}
\begin{center}
\section{Gap function on the Fermi surfaces of the bulk FeSe}
\end{center}

\begin{figure}[htbp]
  \includegraphics[width=1.0\linewidth]{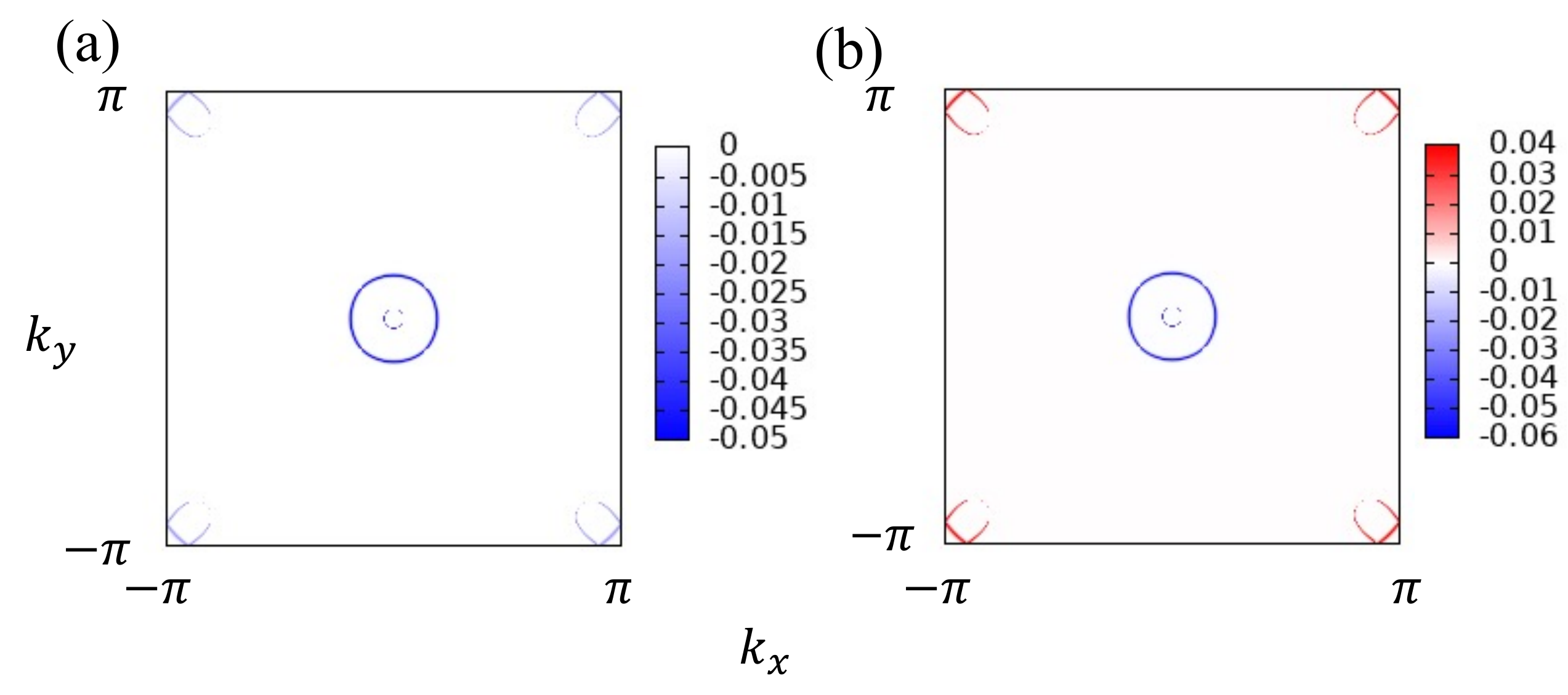}
  \centering
  \caption{
  Gap function on the Fermi surfaces, defined as  $\sum_{m,n} 4T^2\bra{u_{n\bm k}}\bm \Delta\ket{u_{m\bm k}}\times f^\prime(\epsilon^{\rm bulk}_{n\bm k})f^\prime(\epsilon^{\rm bulk}_{m\bm k})$.
  Here, $\epsilon_{n\bm k}^{\rm bulk}$ is the single-particle energy of the bulk FeSe. 
 The attractive interaction is (a) $V_1 = V_2 = 0.2V_0$ and (b) $V_1 = V_2 = 10V_0$.
 From the panels, we find that (a) shows the $s_{++}$-wave pairing state while (b) shows the incipient $s_{+-}$-wave state.}
  \label{fig:gap_fs_kdep}
\end{figure}

Here, we explain the relationship between the gap function and the pairing interaction adopted in this study.
A highly $\bm k$-dependent gap function may show the sign change between the Fermi surfaces of the bulk FeSe near the $\Gamma$ and $M$ points. This gap function corresponds to the incipient $s_{+-}$-wave pairing state in monolayer FeSe because the hole Fermi surfaces vanish owing to the electron doping.
On the other hand, the weakly $\bm k$-dependent gap function does not show the sign change, which is regarded as the $s_{++}$-wave pairing state.

We consider the $\bm k$-dependent pairing interaction,
\begin{eqnarray}
V_{ij\bm k\bm k^\prime} = V_0\delta_{ij}
+ V_1(\delta_{i,j+5} +\delta_{i+5,j})\cos k_x/2\cos k_y/2\cos k_x^\prime/2\cos k_y^\prime/2\notag\\
+ V_2\delta_{ij}(\cos k_x +\cos k_y)(\cos k_x^\prime +\cos k_y^\prime),
\end{eqnarray}
where the pairing interaction on the nearest- and next-nearest-neighbor bonds is taken into account in addition to the on-site pairing interaction.
In Fig.~\ref{fig:gap_fs_kdep}, we plot the gap function on the Fermi surfaces of bulk FeSe.
The parameters for the interaction are (a) $V_1 = V_2 = 0.2V_0$ and (b) $V_1 = V_2 = 10V_0$.
As we see from 
Fig.~\ref{fig:gap_fs_kdep}, the sign of the gap function is the same between the $\Gamma$ and $M$ points in (a), although it is the opposite in (b). 
Thus, the pairing interaction for Fig.~\ref{fig:gap_fs_kdep}~(a) 
leads to the $s_{++}$-wave pairing state, while that for Fig.~\ref{fig:gap_fs_kdep}~(b) 
to the incipient $s_{+-}$-wave pairing state.

\bibliography{suppl}